\documentclass[a4paper,11pt]{article}
\usepackage{pos}
\usepackage{physics}
\usepackage{siunitx}
\usepackage{todonotes}

\title{Axion-like-particle dark matter beyond the standard paradigm}

\author*[a]{Cem Eröncel}

\affiliation[a]{Istanbul Technical University, Department of Physics,\\
  34469 Maslak, Istanbul, Türkiye}

\emailAdd{cem.eroncel@itu.edu.tr}

\abstract{Axions and axion-like particles (ALPs) are among the most popular candidates that explain the origin of the mysterious dark matter. The most popular ALP production mechanism studied in the literature is the misalignment mechanism, where an ALP field with a quadratic or cosine potential has negligible kinetic energy initially, and it starts oscillating when its mass becomes comparable to the Hubble scale. Recently, there has been an interest in models that go beyond the standard assumptions. These models not only extend the ALP dark matter parameter space, but also provide a rich phenomenology which is absent in the standard scenario. In particular, the ALP fluctuations grow exponentially via parametric resonance and tachyonic instabilities. In this proceeding, we will first demonstrate why the standard paradigm cannot explain dark matter in experimentally interesting parts of the parameter space, and then we will give an overview of the alternative production mechanism with which this issue can be resolved. We will then discuss the exponential growth of the fluctuations in these models. Finally, we will comment on the observational consequences of the exponential growth and show that a sizable region of the ALP parameter space becomes testable even if ALPs have only gravitational interactions. }

\FullConference{2nd Training School and General Meeting of the COST Action COSMIC WISPers  (CA21106) (COSMICWISPers2024)\\
 10-14 June 2024 and 3-6 September 2024\\
Ljubljana (Slovenia) and Istanbul (Turkey)\\}

\newcommand{\osc}[1]{#1_{\text{osc}}}

\begin{document}
\maketitle

\section{Introduction}

The QCD axion, or its generalization Axion-like-particle (ALP), is one of the well motivated dark matter candidates. The former also solves the Strong CP problem of the Standard Model. For these reasons, they have been subject to extensive theoretical and experimental research recently. 

ALPs are produced in the early universe mainly as cold relics via the \emph{misalignment mechanism}. In this mechanism, the ALP field is initially misplaced from its vacuum expectation value (VEV) and remains frozen due to the strong Hubble friction. At later times, the Hubble scale decreases and the field starts oscillating around its VEV. The energy density of these oscillations redshifts as pressureless matter so that ALPs can account for all or part of dark matter. 

Recently, there has been a growing interest in going beyond the standard paradigm mentioned above. These extensions not only expand the ALP dark matter parameter space, but they are also theoretically well motivated and produce a rich phenomenology that is absent in the standard scenario. 

In this proceeding, we shall give a succinct review of the recent progress in extending/modifying the misalignment mechanism, which is based on the works~\cite{Eroncel:2022vjg,Eroncel:2022efc,Chatrchyan:2023cmz}. In Sec. \ref{sec:standard} we review the standard paradigm and explain why it is insufficient to produce the correct dark matter density in all parts of the parameter space. Then, in Sec. \ref{sec:extending}, we list possible extensions and discuss how they can expand the parameter space. In Sec. \ref{sec:fluctuations}, we shall see how the fluctuations play a crucial role if one goes beyond the standard paradigm, and in Sec. \ref{sec:observational}, we comment on how the modified mechanism can be probed by observations. Finally, we conclude in Sec. \ref{sec:conclusions}.

\section{Axion-like-particle dark matter in standard paradigm}
\label{sec:standard}

In a cosmological setting, an axion-like-particle (ALP) is represented by a real scalar field which we denote by $\phi$. The cosmological evolution of this field in FLRW spacetime is determined by the differential equation
\begin{equation}
    \label{eq:phi-bg-eom}
    \ddot{\phi}(t,\vb{x})+3H\dot{\phi}(t,\vb{x})-\frac{\nabla^2}{a^2}\phi(t,\vb{x})+V'(\phi)=0.
\end{equation}
In most UV completions, the field $\phi$ arises as the Pseudo-Nambu-Goldstone (pNGB) of a broken global symmetry. In this case, the potential for $\phi$ is generated via non-perturbative effects, and it, in general, depends on the temperature $T$ of the Standard Model plasma:
\begin{equation}
    \label{eq:phi-potential}
    V(\phi,T)=m_{\phi}^2(T)f_{\phi}^2\qty[1-\cos\qty(\frac{\phi}{f_{\phi}})].
\end{equation}
Here, $f_{\phi}$ is the energy scale of symmetry breaking that has generated the ALP as the pNGB, and $m_{\phi}(T)$ is the temperature-dependent ALP mass. For the QCD axion, the latter can be approximated as
\begin{equation}
    \label{eq:mass-dependence}
    m_{\phi}(T)\approx m_{0}\begin{cases}
        (T_c/T)^{\beta},& T\gg T_c\\
        1,&T\ll T_c
    \end{cases},
\end{equation}
with $2\beta\approx 8.16$, and $T_c\sim \sqrt{m_0 f_{\phi}}\approx 75.6\,\si{\mega\electronvolt}$~\cite{Borsanyi:2016ksw}. For a generic ALP, both $\beta$ and $T_c$ are free parameters, and there is no relation between $m_0$ and $f_{\phi}$.


Assuming pre-inflationary scenario and negligible initial kinetic energy, i.e. $\dot{\phi}(t_i)\approx 0$, the evolution of $\phi$ can be divided into two regimes: At early times when $m(T)\ll H(t)$, the field is frozen due to the strong Hubble friction. At late times when $m(T)\gg H(t)$, the field oscillates around the minimum of the potential, $\phi_{\text min}=0$, with a decaying amplitude due to the Hubble friction. In this regime, the comoving number density $a^3 n_{\phi}=a^3 \rho_{\phi}/m_{\phi}$ is conserved. For given set of parameters $\qty{m_0,f_{\phi},\beta,T_c}$, the relic density for ALP dark matter is determined by the initial angle $\theta_i$ and given by the equation~\cite{Arias:2012az}
\begin{equation}
    \label{eq:relic-density-sm}
    h^2 \Omega_{\phi,0}\approx 0.12 \sqrt{\frac{m_{0}}{\si{\electronvolt}}}\sqrt{\frac{m_{0}}{\osc{m}}}\qty(\frac{\osc{m}}{3\osc{H}})^{3/2}\theta_i^2 \qty(\frac{f_{\phi}}{10^{11}\,\si{\giga\electronvolt}})^2 \mathcal{F}(\osc{T}),
\end{equation}
where the subscript "osc" denotes the quantities at the onset of field oscillations, $\mathcal{F}$ is a function of effective degrees of freedom, which varies between $0.3$ and $1.0$. Assuming that the temperature dependence of the ALP mass obeys Eq. \eqref{eq:mass-dependence}, and the ALP field starts oscillating in the regime where its mass is still changing, i.e. $\osc{T}\gg T_c$ we can show that
\begin{equation}
    \label{eq:alp-mass-ratios}
    \sqrt{\frac{m_0}{\osc{m}}}\sim \qty(\frac{\osc{H}}{\osc{m}})^{\frac{\beta/2}{\beta+2}}\qty(\frac{M_{\text{pl}}}{f_{\phi}})^{\frac{\beta/2}{\beta+2}}.
\end{equation}
The case with $\osc{T}\ll T_c$ is indistinguishable from the temperature-independent case, i.e. $\beta=0$.

Eq. \eqref{eq:relic-density-sm} together with Eq. \eqref{eq:alp-mass-ratios} clearly present how the model parameters affect the relic density. From these equations, we can conclude that in the standard paradigm, it is not possible to produce the correct dark matter density if the axion decay constant $f_{\phi}$ is low. To see this, we first note that given the periodic potential in Eq. \eqref{eq:phi-potential}, $\theta_i$ is bounded from above by $\pi$. Furthermore, analytical and numerical calculations show that when $\theta_i$ is not very close to $\pi$, then $\osc{m}/3\osc{H}\sim 1$. Therefore, these two terms cannot compensate for the strong suppression coming from the $f_{\phi}^2$ term. One can argue that an enhancement can be generated by choosing $\beta$ such that $m_0/\osc{m}$ is large. But this is not effective since for $\osc{m}\sim\osc{H}$ we find
\begin{equation}
    \label{eq:f_dependence}
    h^2\Omega_{\phi,0}\sim \qty(\frac{f_{\phi}}{M_{\rm pl}})^{2-\frac{\beta/2}{\beta+2}}\xrightarrow{\beta\rightarrow \infty} \qty(\frac{f_{\phi}}{M_{\rm pl}})^{3/2}.
\end{equation}
So, even a large value for $\beta$ does not help to obtain the correct the relic density if $f_{\phi}$ is low. 

This discussion shows that we need to go beyond the standard paradigm to extend the ALP dark matter parameter space to lower $f_{\phi}$ values. Such an extension is desirable since low $f_{\phi}$ region is experimentally more interesting due to the fact ALPs do couple more strongly to the Standard Model.

\section{Extending the ALP dark matter parameter space}
\label{sec:extending}

Let us discuss how we can go beyond the standard paradigm and thereby extend the ALP dark matter parameter space to lower values of the axion decay constant. The challenge is to enhance the dark matter abundance so that sufficient amount of dark matter can be produced even for lower values of $f_{\phi}$. By inspecting Eq. \eqref{eq:relic-density-sm}, we can see that there are two possibilities. We can either delay the onset of oscillations such that $\osc{m}/\osc{H}\gg 1$ or we can modify the periodic potential given in Eq. \eqref{eq:phi-potential} to a non-periodic one such that the bound $\theta_i^2<\pi^2$ does not apply. In this proceeding, we will focus on the former case, since it can also be applied to the QCD axion. The discussion of the latter case can be found in~\cite{Chatrchyan:2023cmz,Chatrchyan:2024vab}.

The delay of the onset of oscillations can be achieved by either choosing the initial angle $\theta_i$ to be very close to the maximum of the periodic potential, i.e. $\abs{\pi-\theta_i}\ll 1$, or by assuming that the ALP field has a large initial kinetic energy. The former case is known as the \emph{Large Misalignment Mechanism (LMM)}~\cite{Arvanitaki:2019rax} or \emph{extreme axion}~\cite{Zhang:2017dpp} in the literature. In this case, the onset of oscillations is delayed due to the tiny gradient at the top of the potential, which can be approximated by~\cite{Arvanitaki:2019rax}
\begin{equation}
    \label{eq:lmm-enhancement}
    \frac{\osc{m}}{\osc{H}}_{\text{LMM}}\sim \ln\qty[\frac{1}{\pi-\abs{\theta_i}}\frac{2^{1/4}\pi^{1/2}}{\Gamma(5/4)}].
\end{equation}
However, due to the logarithmic dependence, large tunings are required for significant enhancements. Therefore, LMM can extend the parameter space by only about one order of magnitude~\cite{Eroncel:2022vjg}. 

The model where the ALP field has a large kinetic energy is known as the \emph{Kinetic Misalignment Mechanism (KMM)}~\cite{Co:2019jts,Chang:2019tvx}. Such a kinetic energy can be motivated in various ways, for example by explicit breaking of the Peccei-Quinn symmetry at high energies~\cite{Co:2019jts,Co:2020dya,Co:2020jtv} or by considering a non-trivial temperature dependence for the ALP potential~\cite{DiLuzio:2021gos,DiLuzio:2021pxd}. In KMM, at early times the ALP field is kinetic energy dominated, which redshifts as $\rho_{\phi}\propto a^{-6}$. The oscillations start once the kinetic energy becomes comparable to the potential energy. For a given ALP mass $m_0$, to generate sufficient amount of dark matter for smaller $f_{\phi}$ one needs a larger $\osc{m}/\osc{H}$. However there is no a priori constraint against this\footnote{For an extensive discussion of which part of the ALP dark matter parameter space can be accessed by realistic UV completions, we refer the reader to~\cite{Eroncel:2024rpe}.}. 

\section{Fluctuations of the ALP field}
\label{sec:fluctuations}

Even in the pre-inflationary scenario, the ALP field has some fluctuations on top of the homogeneous background.  These fluctuations are seeded by adiabatic and isocurvature perturbations. The former are due to the energy density perturbations of the dominant component in the universe, and they are unavoidable.
In this proceeding, we will only consider adiabatic fluctuations.

By expanding the ALP field as $\phi(t,\vb{x})=\bar{\phi}(t)+\delta \phi(t,\vb{x})$ where $\bar{\phi}$ and $\delta \phi$ denote the background and fluctuations respectively, we can derive the following equation of motion for the Fourier modes $\phi_k$ of the fluctuations at linear order in perturbations:
\begin{equation}
    \ddot{\phi}_k+3H\dot{\phi}_k + \qty[\frac{k^2}{a^2}+\eval{V''(\phi)}_{\bar{\phi}}]\phi_k=2\Phi_k \eval{V'(\phi)}_{\bar{\phi}}-4\dot{\Phi}_k\dot{\bar{\phi}}.
\end{equation}
Here $k$ is the comoving momentum of the Fourier mode, and $\Phi_k$'s are the Fourier modes of the curvature fluctuations. This equation of motion is unstable when the effective frequency $k^2/a^2 + \eval{V''(\phi)}_{\bar{\phi}}$ becomes negative and/or is oscillating~\cite{Kofman:1997yn,Felder:2006cc,Greene:1998pb,Jaeckel:2016qjp,Cedeno:2017sou,Berges:2019dgr,Fonseca:2019ypl,Morgante:2021bks}. For the ALP potential given in Eq. \eqref{eq:phi-potential} both kinds of instabilities exist. In fact, instabilities exist for every potential except the potential of a free theory where $V''(\phi)=m^2$. Due to instabilities, some Fourier modes experience exponential growth during their evolution. This effect is studied extensively in~\cite{Arvanitaki:2019rax} for the LMM case and in~\cite{Eroncel:2022vjg,Eroncel:2022efc} for the KMM case. There, it has been demonstrated that the growth rate of the perturbations depends exponentially on $\osc{m}/\osc{H}$. The reason for this dependence can be understood as follows: When $\osc{m}\sim\osc{H}\sim 1$, Hubble quickly dampens the field amplitude; therefore, after a few oscillations the ALP field effectively evolves in a quadratic potential without any instability. However, when $\osc{m}\gg\osc{H}$ the field amplitude decreases much more slowly, so that fluctuations experience instabilities for a much longer time. 

When $\osc{m}/\osc{H}$ reaches to a critical value, which is found to be $\sim 40$ in the case of KMM~\cite{Eroncel:2022vjg}, the exponential growth becomes so strong that the power spectrum reaches to $\mathcal{O}(1)$ values. After this, the linear perturbation theory breaks down, and the ALP field cannot be considered a homogeneous field anymore. For the KMM scenario, this regime is studied semi-analytically via an energy conservation argument in~\cite{Eroncel:2022vjg}. There it has been found that the non-linear effects smoothen out the power spectrum, so that in this regime stronger exponential growth yields a power spectrum with smaller peaks. This result has also been confirmed in~\cite{Chatrchyan:2023cmz} via lattice simulations, albeit for a non-periodic potential. 

\section{Observational Consequences}
\label{sec:observational}
The most important consequence of the exponential growth of the ALP fluctuations is that dense and compact ALP mini-clusters, which are usually considered smoking gun signatures of the post-inflationary scenario can also be formed in the pre-inflationary scenario. The resulting mini-cluster spectra have been calculated semi-analytically using the Excursion Set Formalism in~\cite{Arvanitaki:2019rax},~\cite{Eroncel:2022efc}, and~\cite{Chatrchyan:2023cmz} for LMM, KMM, and non-periodic potentials, respectively. There, it has been found that for a given ALP mass $m_{0}$ and a production mechanism such as LMM, KMM, non-periodic, etc., there is a critical $f_{\phi}$ that yields the most dense structures. Based on this observation, Ref.~\cite{Chatrchyan:2023cmz} derives a "dense halo region" in the ALP dark matter parameter space, which indicates the parameter space where the exponential growth might yield dense structures that are likely to survive tidal stripping. 

These dense halo regions in different production mechanisms mostly overlap with each other. So, it is difficult to infer the mechanism from observations. However, observations of dense structures provides us with information about the ALP mass and the decay constant even when ALP does not couple to the Standard Model. 

\section{Conclusions}

We can summarize the main points of this review as follows: The Standard Misalignment Mechanism is not sufficient to account for the correct dark matter abundance in the ALP parameter space, where the experiments are most sensitive. This parameter space can be opened by considering models where the initial energy budget is increased and the onset of oscillations is delayed from the conventional value $\osc{m}/\osc{H}\sim 3$. In these models, which go beyond the standard paradigm, the fluctuations can grow exponentially, and dense mini-clusters can be formed even in the pre-inflationary scenario. Semi-analytical studies predict that there is a band on the $(m_{\phi},f_{\phi})$-plane where dense structures can be formed, and the location of this band does not depend drastically on the production mechanism. The existence of this band allows us to obtain information about the decay constant, even if ALP does not couple to the Standard Model. 
\label{sec:conclusions}

\section*{Acknowledgements}
This article/publication is based upon work from COST Action COSMIC WISPers CA21106, supported by COST (European Cooperation in Science and Technology).
\bibliographystyle{JHEP}
\bibliography{references}

\end{document}